\documentclass[12pt]{article}

\usepackage[usenames,dvipsnames,svgnames,table]{xcolor}

\usepackage{amsmath, amssymb}
\usepackage{bm, booktabs}
\usepackage{cancel, caption, color}
\usepackage{dcolumn}% Align table columns on decimal point
\usepackage{epsfig, epsf, enumitem}
\usepackage{fancyhdr}
\usepackage[T1]{fontenc}
\usepackage{graphicx, geometry}
\usepackage{hyperref}
\usepackage{ifthen}
\usepackage[utf8]{inputenc}
\usepackage{lscape, longtable}
\usepackage{multirow}
\usepackage{natbib}
\usepackage{pifont}
\usepackage{ragged2e}
\usepackage{subfigure}
\usepackage{sectsty}
\chapterfont{\color{blue}}                   
\sectionfont{\color{cyan}}                   
\subsectionfont{\color{ProcessBlue}}  
\usepackage{times, tabularx}
\usepackage{tcolorbox}
\usepackage{verbatim}
\tcbuselibrary{skins}
\newcolumntype{Y}{>{\raggedleft\arraybackslash}X}

\tcbset{tab1/.style={enhanced, fonttitle=\bfseries, fontupper=\normalsize\sffamily,
colback=yellow!10!white,
colframe=red!50!black,
colbacktitle=Cerulean!40!white,
coltitle=black,center title}
}

\setitemize{noitemsep,topsep=0pt,parsep=0pt,partopsep=0pt,leftmargin=*}
\def\jcap{\ref@jnl{J. Cosmology Astropart. Phys.}}%

\def \gsim { \lower .75ex \hbox{$\sim$} \llap{\raise .27ex \hbox{$>$}} }
\def \lsim { \lower .75ex \hbox{$\sim$} \llap{\raise .27ex \hbox{$<$}} }

\begin{document}
\raggedright
\huge
Astro2020 Science White Paper \linebreak

\textcolor{Cerulean}{Opportunities in Time-Domain Extragalactic Astrophysics with the NASA
Near-Earth Object Camera (NEOCam)}
\normalsize
\linebreak \linebreak 

\noindent \textbf{Thematic Areas:} \\
Primary:: Galaxy Evolution  \\
Secondary:: Cosmology and Fundamental Physics/Multi-Messenger Astronomy and Astrophysics 
\linebreak
  
\textbf{Principal Author:}

Name: Nicholas P. Ross	
 \linebreak						
Institution:  University of Edinburgh
 \linebreak
Email: npross@roe.ac.uk
 \linebreak
Phone:  +44 (0)131-668 8351
 \linebreak
 
\textbf{Co-authors:} %% (names and institutions)
  \linebreak
Roberto J. Assef (Universidad Diego Portales)   \linebreak
Matthew J. Graham (Caltech) \linebreak 
J. Davy Kirkpatrick (Caltech/IPAC) \linebreak

\textbf{Endorsers:} %% (names and institutions)
 \linebreak
Aaron M. Meisner (NOAO) 
\linebreak

\justify
\textbf{Abstract:}
This White
Paper\footnote{Which can be found, with supporting documents at \href{https://github.com/d80b2t/Astro2020}{{\tt github.com/d80b2t/Astro2020}}}
motivates the {\it time domain extragalactic science case} for the
NASA Near-Earth Object Camera (NEOCam). NEOCam is a NASA Planetary
mission whose goal is to discover and characterize asteroids and
comets, to assess the hazard to Earth from near-Earth objects, and to
study the origin, evolution, and fate of asteroids and comets. NEOCam
will, however, cover 68\% of the `extragalactic sky' and as the
NEOWISE-R mission has recently proved, infrared information is now
vital for identifying and characterizing the $\gtrsim$10 million IR
bright Active Galactic Nuclei, as well as using the IR light curve to
provide deep insights into accretion disk astrophysics. NEOWISE-R data
has also been used to discover Super-luminous Supernovae, dust echos
in Tidal Disruption Events and detects all of the known $z\geq7$
quasars (and over 80\% of the known $z\geq6.70$ quasars). As such, for
relatively little additional cost, adding the capacity for additional
NEOCam data processing (and/or alerting) would have a massive
scientific and legacy impact on extragalactic time domain science.

\pagebreak
\smallskip
\smallskip
\noindent
{\bfseries \textsc{\textcolor{Cerulean}{Time Domain Extragalactic Science Now and into the Next Decade}}}

\smallskip
\smallskip 
\noindent
The field of time domain extragalactic science started at least in the 2000s, with Palomar-Quest, LINEAR, CRTS, OGLE, EROS and MACHO. The field then moved into the 2010s with the SDSS Stripe 82, Pan-STARRS, PTF and ZTF, and with the arrival of LSST, time domain extragalactic science will be a fully mature field in the 2020s. However, the {\it mid-infrared extragalactic time domain field} was completely unexplored until the {\it Spitzer} Deep Wide-Field Survey (Koz{\l}owski et al. 2010, 2016), which has now become the Decadal IRAC Bo\"{o}tes Survey (Ashby et al. 2013). The advent of the Wide-field Infrared Survey Explorer mission and its asteroid-characterization extension, the Near-Earth Object Wide-field Infrared Survey Explorer Reactivation mission, fully opened this new window.

\smallskip 
\smallskip 
\noindent
The Wide-field Infrared Survey Explorer (WISE; Wright et al. 2010) is
a NASA infrared (IR) wavelength astronomical space telescope launched
in December 2009, which performed a ``Cold'' mission imaging out to
28$\mu$m in four bands centered at wavelengths of 3.4, 4.6, 12, and
23$\mu$m. The WISE satellite then performed an extended mission called
Near-Earth Object WISE (NEOWISE; Mainzer et al. 2011) though only in
the two shortest wavelength bands. The WISE satellite was then placed in
hibernation in February 2011, but was reactivated in December 2013.
Since then, the NEOWISE-Reactivation (NEOWISE-R; Mainzer et al. 2014)
mission has been mapping the full sky. NEOWISE-R survey observations
are continuing in 2019, and as of March 2019, NEOWISE is over 40\%
of the way through its 11th coverage of the sky since the start of the
Reactivation mission.

\smallskip
\smallskip 
\noindent
Critically, these 11 epochs of 3-5$\mu$m infrared data have now been
used in conjunction with active galactic nuclei (AGN) and quasar
datasets in order to find large numbers of dramatically varying
quasars for the very first time (e.g., Assef et al. 2018a, 2018b; Stern
et al., 2018; Ross et al. 2018, Yan et al. 2019). NEOWISE data has
also been critical in detecting dust echos associated with Tidal
Disruption Events (TDEs; e.g., Dou et al., 2016, 2017; French et al., 2018;
Jiang et al., 2016, 2017; Wang et al. 2018). 

\smallskip
\smallskip
\noindent
Because NEOWISE-R is an asteroid discovery and characterization, the mission itself does not publish any coadded data product of the sort that maximize the raw NEOWISE data’s value for extragalactic (time-domain) astrophysics. However, building on the initial work of Lang, Schlegel and Hogg (e.g. Lang 2014; Lang et al. 2016), Meisner, Schlafly and Green have lead a wide-ranging effort to repurpose NEOWISE observations for astrophysics, starting by building deep full-sky coadds from the vast publicly available archive of WISE and NEOWISE exposures (e.g. Schlafly  et al., 2019, ApJS, 240, 30 and references therein).

\smallskip 
\smallskip
\noindent
The resulting ``unWISE'' line of data products\footnote{The ``un''
refers to images having {\it not} been re-convolved with the PSF
before the coaddition, resulting in better resolution than the ALLWISE
data products; see \href{http://unwise.me/}{{\tt http://unwise.me/}}.}
have already created the deepest ever full-sky maps at 3–5$\mu$m
(Meisner et al. 2017a,b,c; Meisner et al. 2018a), generated a new
class of time-domain WISE coadds (Meisner et al.  2018b,c), and
performed forced photometry on these custom WISE coadds at the
locations of more than a billion optically detected sources (Lang et
al. 2014; Lang et al. 2016; Dey et al. 2018, Schlafly et
al. 2019). The NEOWISE/unWISE data are also a critical part of the
target selection for the DOE/NSF/NOAO Dark Energy Spectroscopic
Instrument (DESI) and cosmology survey\footnote{\href{https://www.desi.lbl.gov/}{{\tt https://www.desi.lbl.gov/}}}.

\smallskip
\smallskip 
\noindent
The NASA Near-Earth Object Camera
(NEOCam\footnote{\href{https://neocam.ipac.caltech.edu/}{{\tt https://neocam.ipac.caltech.edu/}})}
is a NASA Planetary mission, currently in Extended Phase A, whose goal is to
discover and characterize asteroids and comets, carrying on the legacy
of NEOWISE. {\it The main purpose of this White Paper is to encourage,
in the strongest terms, NASA Astrophysics to budget and plan for a
line of e.g. coadded data products from the NEOCam mission that will
result in a substantial extragalactic resource. As NEOWISE-R and
unWISE have proved, the ``Return On Investment'' here is
considerable.}

\smallskip
\smallskip
\noindent
{\bfseries \textsc{\textcolor{Cerulean}{The NASA Near-Earth Object Camera (NEOCam)}}}

%\smallskip
\smallskip
\noindent
\textsl{\textsc{Overview:}}
NEOCam is a single-instrument, 50cm diameter telescope that will observe in two
simultaneous channels, called NC1 and NC2, which cover wavelengths of
4.0-5.2 $\mu$m and 6.0-10.0$\mu$m, respectively. There are four Teledyne
2k$\times$2K HgCdTe arrays per channel with 3.00'' pixels in each. The entire
celestial sphere between ecliptic latitudes of +40$^{\circ}$ to -40$^{\circ}$ will
continually be mapped over the course of the 5-yr mission, and over 12
yr for the design lifetime. 

\smallskip
\smallskip
\noindent
{\it 
NEOCam would survey the Solar System at two simultaneous thermal IR
bandpasses and would also provide a synoptic survey of two-thirds of
the thermal infrared sky at 4 - 5.2$\mu$m and 6 - 10$\mu$m using a
large field of view of 11.56 square degrees.
%NEOCam will provide a synoptic survey of two-thirds of the thermal infrared sky at 4-5.2 microns and 6-10 microns.
}

\smallskip
\smallskip
\noindent
\textsl{\textsc{Cadence:}} Estimates of on-board overheads and resulting orbit quality are still being refined, but the current data cadence is as follows. The observing pattern consists of a four-peat of a six-position dither (a quick sequence of six images with 28s integration time each) with 2h gaps between each repeat. This four-peat will recur $\sim$13.2d later as long as the 75d visibility window is still open. Afterward, there will be a gap of 215d until the next visibility window opens and the pattern begins again. On average in a typical visibility window, there are $\sim$23 of these six-position dither sequences (a little less than 6 four-peats), $\sim$234 over 5yr, and $\sim$562 over 12yr. In addition, observations covering calibration targets with well-known spectra and photometry are expected to be taken periodically to establish zero points and monitor the instrument's photometric stability. 

\smallskip
\smallskip
\noindent
\textsl{\textsc{Depth:}} Each of the six-position dither sequences is required to have an S/N=5 sensitivity of 65-120 $\mu$Jy for NC1 and 110-280 $\mu$Jy for NC2, for low to high zodiacal backgrounds. (For comparison WISE achieved 80, 110 and 1000 $\mu$Jy 5$\sigma$ point source sensitivities in the $\lambda=2.8-3.8$, $4.0-5.2$ and $7.5-17\mu$m W1, W2 and W3 bands, respectively).

\smallskip
\smallskip
\noindent
\textsl{\textsc{Data Products:}} NEOCam processing will create images and lists of characterized sources from each individual exposure and each stacked six-dither position sequence. It will also create differenced images by subtracting a static reference image, and a list of the characterized transient detections also produced. Because the goal of NEOCam is to provide and characterize moving objects within the solar system, coadd and source extractions over longer timescales are not provided, the one exception being yearly builds to create new static images (without any source detection or characterization) of the sky to use in image differencing. No alerting mechanism is provided for astrophysical transient events.  

\smallskip
\smallskip
\noindent
{\bfseries \textsc{\textcolor{Cerulean}{Continuing the NASA IR Space Telescope Legacy}}} 

\smallskip
\noindent
The Infrared Astronomical Satellite (IRAS; 1983 January 25th launch, Wheelock et al. 1994), the {\it Spitzer} Space Telescope (SSC; 2003 August 25th launch) and the Wide-field Infrared Survey Explorer (WISE; 2009 December 14th launch) satellites all show how powerful infrared space telescopes, with moderate-sized primary mirrors are. We do also note, that although a top NASA priority, the {\it JWST} variability programs are restricted to very small regions (e.g., Jansen \&  Windhorst, 2018). The recently selected SPHEREx mission (Dor{\'e} et al. 2016, 2018) will be an excellent near-IR complement to NEOCam, obtaining spectra over 0.75-5$\mu$m across the full sky, but SPHEREx is not as deep as WISE, and currently will only map the entire sky four times during its nominal 25-month mission. Thus, NEOWISE is essentially the key and only {\it mid-infrared time domain extragalactic} dataset we have available until NEOCam comes online.

\smallskip
\smallskip
\noindent
{\bf To realize the full potential of the NEOCam data for
astrophysical research, additional data products and alerting
infrastructure are needed. For a relatively small investment, NASA can
leverage the existing NEOCam data to enable a wide range of
extragalactic, time-domain, research.}

\smallskip
\smallskip
\noindent
%In particular, extragalactic science has always benefitted from space-based IR operations, and this will continue in the 2020s with JWST, as well as new very wide-field observatories such as NEOCam. 
%%
{\it We envisage that by releasing and maintaining high-quality, publicly available archives of NEOCam data -- 
in order to enable high-ROI archival data analyses -- 
NEOCam will have a direct and considerable impact on the NASA Astrophysics Division {\it Physics of the Cosmos} and {\it Cosmic Origins} programs.}
Here we highlight particular extragalactic science cases all of which have either an infrared time-domain aspect, or that have benefitted from NEOWISE-R archival analyses extending beyond the nominal data products.

\smallskip
\smallskip
\noindent
{\bfseries \textsc{\textcolor{Cerulean}{infrared Variability: A
cornerstone of AGN and Quasar investigations in the next Decade and
beyond}}}

\smallskip
\smallskip
\noindent
In AGN, circumnuclear dust absorbs the accretion disk illumination and
reradiates the absorbed energy in the IR (e.g. Li 2007). The geometry
of the dusty obscuring medium is not known, but depending upon whether or 
not the dust covers the accretion disk, we will observe the AGN as being optically obscured (`type 2') or unobscured
(`type 1'). The IR emission at wavelengths longward of $\lambda > 1
\mu$m accounts for at least 50\% of the bolometric luminosity of type
2 AGNs. For type 1 AGNs, $\sim$10\% of the bolometric luminosity is
emitted in the IR (see e.g. Osterbrock \& Ferland 2006; Li 2007). A
near-IR ``bump'' (excess emission above the $\sim$2–10$\mu$m
continuum), is generally attributed to hot dust with temperatures
around $\sim$1200–1500K (near the sublimation temperatures of
silicate and graphite grains), and is seen in a few type 1 AGNs
(Barvainis 1987; Rodriguez-Ardila \& Mazzalay 2006).
%% TORUS (The Obscuring Region of the Unified Scheme)

\smallskip
\smallskip
\noindent
As discussed in detail in Stern et al. (2005) and Lacy et al. (2004), the lack of strong
polycyclic aromatic hydrocarbon emission in powerful AGNs, along with
the IR flux $\lambda_{\rm rest} < 5 \mu$m flux of AGNs being dominated
by power-law emission rather than a composite stellar spectrum, leads
to the AGNs being significantly redder than lower-redshift galaxies.

\smallskip
\smallskip
\noindent
Even though we can now readily identify AGN from their IR signatures, 
there are several key outstanding questions:  

\begin{itemize}
\item Do we have a full accounting for the bolometric energy in AGN?
\item What is the geometry of the infrared emitting region of a AGN central engine?
\item How does the (UV/optical) emission from the inner accretion disk impact on the infrared emitting region?
\item What does IR variability tell us about black hole mass build up?
\item What is the deep connection between high-energy neutrinos and AGN blazars? 
\end{itemize}

\noindent
{\it In summary, what will the variable IR emission tell us about AGN energetics and the 
role of black holes in galaxy formation?}

\smallskip
\smallskip
\noindent
\textbf{\textsc{Super-Luminous Supernova in AGN torus: }} 
%\smallskip \smallskip \noindent
Assef et al. (2013, 2018a; also see Stern et al. 2012) have led the field in identifying large samples of AGN candidates selected from WISE (for a review of mid-IR AGN selection see Padovani et al. 2017). From their most reliable sample of AGN candidates, Assef et al. (2018a) presented a sub-sample of 45 objects with the strongest level of IR variability in WISE but with no radio detection in the FIRST survey, implying that they are not blazars. Assef et al. (2018b) presented a detailed study of these 45 objects and found that only seven of them showed significant co-eval optical variability in the Catalina Real-time Transient Survey (CRTS; Drake et al., 2009). 

\smallskip
\smallskip
\noindent
Further investigation revealed that in one of these seven optically
variable objects, W0948+0318, the optical and IR variability is most
likely driven by a super-luminous supernova (SLSN) with a total
radiated energy of $E=1.6\pm 0.3 \times 10^{52}~\rm erg$, which makes
it one of the most energetic SLSNe ever observed (see Assef et
al. 2018b for details). Based on the lack of change in the mid-IR
color throughout the transient, they speculate that the SLSN occurred
within the dust structure of the AGN. Additionally, Assef et
al. (2018b) identified nine possible analogs to W0948+0318 based on
their WISE lightcurves, implying that if they share a similar nature,
they are heavily obscured and could not be identifiable without mid-IR
time-domain surveys. As the depth and cadence of NEOCam will greatly
exceed those of WISE, NEOCam is posed to identify a large numbers of
similar, possibly even more energetic, transients.

\smallskip
\smallskip
\noindent
\textbf{\textsc{AGN and quasar central Engines via the ``Changing Look'' phenomenon: }} 
%\smallskip \smallskip \noindent
In ``Changing-Look quasars'' (CLQs), the strong UV continuum and broad hydrogen emission either appear or disappear on observed-frame timescales of months-to-years (e.g., LaMassa et al. 2015; Macleod et al. 2016; Ruan et al. 2016a, 2016b; Runnoe et al. 2016; Gezari et al. 2017; Yang et al. 2018; Stern et al. 2018; Ross et al. 2018; Parker et al. 2019). Given the luminosity of these objects and the structure function, these changes are unexpectedly fast. These timescales can potentially be associated with the viscous timescale (drift time through the accretion disk), the light crossing timescale (critical for reverberation mapping and disk reprocessing) and the dynamical timescale of the system.  {\it CLQs are thus an ideal laboratory for studying accretion physics, as the entire system responds to a large change in ionizing flux on a human timescale}. The physical processes responsible for these changing-look quasars are still debated, but physical changes in the accretion disk structure appear to be the more likely cause rather than changes in obscuration. However, this leads to a irreconcilable breakdown of the classical, thin accretion disk model, leading to what has recently been called the ``Quasar Viscosity Crisis'' (Lawrence, 2018). 

\smallskip
\smallskip
\noindent
%\textbf{\textsc{New IR investigations into the CLQ Population:}}
Taking advantage of deeper, longer time baseline 
IR light-curves from NEOWISE-R (Meisner et al., 2017a,b), we have begun to make in-roads
into understanding the CLQ population.  This includes identifying
objects with rapidly changing IR light-curves and also accretion disk
changes, e.g. the $z=0.378$ quasar SDSS J1100-0053 (Ross et al., 2018)
and WISE J1052+1519 (Stern et al. 2018). Using these quasars as
archetypes a new model (e.g. Stern et al., 2018, Ross et al., 2018,
Ford et al. 2019) suggests a dramatic new picture of the physics of
the CLQs governed by processes and structure of the very innermost disk.

\smallskip
\smallskip
\noindent
Without any changes to the scientific requirements, or mission profile, NEOCam offers 
a dramatic new dataset for extragalactic time domain astrophysics investigations, including
the long term monitoring of the CLQ population that we are identifying today.
{\it With infrared emission from AGN associated with structures $\sim$a few light months to $\sim$several light years from the central engine accretion disk and photon source, the variable AGN observed in the UV/optical during the early/mid 2020s from LSST will be NEOCam extragalactic time domain sources.}

\smallskip
\smallskip
\noindent
\textbf{\textsc{Infrared Flares as Dust Echos in Tidal Disruption Event candidates:}}   
The sporadic accretion following the tidal disruption event (TDE) of a star by a super-massive black hole leads to a bright UV and soft X-ray flare in the galactic nucleus. The gas and dust surrounding the black hole responds to such a flare with an echo in emission lines and infrared emission. This mid-infrared echo from the TDE candidates is now readily observed with NEOWISE (e.g., van Velzen et al., 2016, Dou et al.,  2016, 2017;  Jiang et al., 2016, 2017, 2019; Wang et al., 2018; Decker et al. 2018). {\it Interpreting these flaring IR signatures will lead to 
a deeper understanding of the physics of accretion (via a disk or stellar 
material source), the dusty central engine geometry and the energy
outsourced from the AGN to the host galaxy.}

\smallskip
\smallskip
\noindent
\textbf{\textsc{Super-critical Accretion in Very High redshift Quasars:}} 
Very high redshift quasars are excellent probes of the early Universe. This includes studies of the Epoch of Reionization for hydrogen (see e.g., Fan et al., 2006; Mortlock 2016 for reviews), the formation and build-up of supermassive black holes (e.g., Rees 1984; Wyithe \& Loeb2003; Volonteri2010; Agarwal et al., 2016; Valiante et al., 2018; Latif et al, 2018; Wise et al., 2019) and early metal enrichment (see e.g., Simcoe et al. 2012, Chen et al. 2017, Bosman et al. 2017).

\smallskip
\smallskip
\noindent
Recently, Ross \& Cross (2019) have compiled a database of all spectroscopically confirmed $z\geq5.00$ quasars, totalling 463 objects. Of these, 283 are detected in the WISE ``ALLWISE'' W1/2 data release. However, this detection rate increases to 362 in W1 and 308 in W2 in the unWISE catalogs, i.e. an increase in detection rate by 28\% and 9\%, in W1 and W2, respectively (the equivalent to a 5.00 year mission being run for 6.4 and 5.45 years). Arguably even more impressive, is that 17  out of  21 (81\%) of the $z\geq6.70$ quasars and all of the four known $z\geq7.00$ quasars are detected in the unWISE catalog. These authors looked for variability as a signature of super-critical accretion in this sample, but with a sampling of only $\lesssim1$year in the quasar rest-frame, no obvious examples were found. 
{\it Increasing the NEOWISE time baseline with NEOCam to several years/a decade will be 
critical for the study of very high-$z$ quasar physics.} 
% \smallskip \smallskip\noindent
%{\it 
%}

\smallskip
\smallskip
\noindent
\textbf{\textsc{Connecting AGN activity in the IR with multi-messenger blazars: }}
With the identification of an electromagnetic counterpart to GW1708017 (e.g., Abbott et
al. 2017a,b,c, Cowperthwaite et al. 2017, Soares-Santos, et
al. 2017)
and the neutrino emission from the direction of the blazar TXS 0506+056 
(IceCube Collaboration; 2018a,b).
we have  fully entered the era of ``multi-messenger astronomy''. 

\smallskip
\smallskip
\noindent
Massaro et al. (2011) identified the WISE Blazar Strip (WBS), a region in WISE [3.4]–[4.6]–[12] $\mu$m color–color space where sources dominated by thermal radiation are separated from those dominated by non-thermal emission, i.e. from the blazar population. D'Abrusco et al. (2012) and Mao et al. (2018) then studied the MIR variability from WISE and NEOWISE-R, respectively, of a large sample of blazars, that had both detections and non-detections by the {\it Fermi}-LAT. These authors found that first, {\it Fermi}-detected blazars tended to be more variable (in the IR) than non-{\it Fermi}-detected blazars, and second, for the {\it Fermi}-detected blazars, there are highly significant correlations between the flux densities and spectral indices in the MIR and gamma-ray bands. These results imply that the activity in the MIR and gamma-rays is connected. 
{\it Thus, we now have a direct connection between extragalactic infrared time domain and neutrino messenger astrophysics.}

%With the 3rd Generation Ground-based Gravitational-wave Observatory Network (``3G'') and the ESA-NASA Laser Interferometer Space Antenna (LISA). 

%\hspace{-78pt}
%
%\smallskip
%\smallskip
%\noindent
%\input{KeyNumbers_table}
%
%
\pagebreak
\noindent
\textbf{References} \\
%%
%\begin{center}
%\medskip
% \medskip
% {\large \bf References}
%    \vspace{-10pt}
%\end{center}
%\begin{multicols}{3}[]
%\noindent
%%\footnotesize
%\scriptsize
%%\tiny
%%
Abbott et al., 2017a,  PhRvL, 119p1101A	\\
Abbott et al., 2017b, ApJ, 848, L12	\\
Abbott et al., 2017c, ApJ, 848, L13	\\
Agarwal et al., 2016,  MNRAS, 459, 4209 \\
Ashby et al., 2013, sptz.prop10088A \\
Assef et al., 2013, ApJ, 772, 26      \\       %% MIR WISE Selection Paper II
Assef et al., 2015,  ApJ, 804,27 \\            %% Half of the quasars are Hot DOGs 
Assef et al., 2018a,  ApJS, 234, 23	\\     %% WISE AGN Catalog
Assef et al., 2018b, ApJ, 866, 26 \\          %% SLSN 
Barvainis, 1987, ApJ, 320, 537 \\
Bosman et al., 2017, MNRAS, 470, 1919 \\
Chapman \& Shanks, 2019, MNRAS, {\it in prep.} \\
Chen et al., 2017, ApJ, 850, 188 \\
Cowperthwaite et al., 2017,  ApJ, 848, L17	\\
Decker, 2018, ApJ, 868, 99 \\	
Dey et al. 2018, AJ accepted, arXiv:1804.08657v2 \\
Dor{\'e} et al., 2016, arXiv:1606.07039v1\\
Dor{\'e} et al., 2018, arXiv:1805.05489v2 \\
Dou et al., 2016, ApJ, 832, 188 \\ 
Dou et al., 2017, ApJL, 841, L8 \\
Drake et al. 2009, ApJ, 696, 870 \\
Fan, Carilli \& Keating, 2006, ARA\&A, 44, 415 \\
French et al., 2018, ApJ, 868, 99 \\
IceCube Collaboration, et al., 2018, Science, 361, 146 \\ 
IceCube Collaboration, et al., 2018, Science, 361, 147 \\
Jansen \& Windhorst, 2018, PASP, 130l4001J \\
Jiang et al., 2016, ApJL, 828, L14 \\
Jiang et al., 2017, ApJL, 841, L8\\ 	
Jiang et al., 2017, ApJ, 850, 63  \\
Jiang et al., 2019, ApJ, 871, 15 \\
Koz{\l}owski et al., 2010, ApJ, 716, 530 \\  
Koz{\l}owski et al., 2016, ApJ, 817, 119 \\
Lacy et al., 2004, ApJS, 154, 166 \\
Lawrence, 2018, Nature Astronomy, 2, 102 \\
Latif et al., 2018, arXiv:1801.07685v1 \\
Lang, 2014, AJ, 147, 108	\\
Lang, Hogg, \& Schlegel, 2016, AJ, 151, 36 \\
Li, 2007, ASPC, 373, 561\\
Mainzer et al., 2011, ApJ, 731, 53 \\
Mainzer et al., 2014, ApJ, 792, 30 \\
Mao et al., 2018,  Ap\&SS, 363, 167\\   %% Mid-infrared variability of blazars: a view from NEOWISE survey
Massaro et al., 2011, ApJ, 740, L48 \\ %% Identification of the Infrared Non-thermal Emission in Blazars
Meisner et al., 2017a, AJ, 153, 38 \\    
Meisner et al., 2017b, AJ, 154, 161 \\
Meisner et al., 2018a, RNAAS, 2, 1 \\
Meisner et al., 2018b, AJ, 156, 69 \\
Meisner et al., 2018c, RNAAS, 2, 202 \\
Mortlock, 2016, ASSL, 423, 187 \\
Osterbrock \& Ferland, 2006, {\it Astrophysics of gaseous nebulae and AGN}   \\ %% Fig 13.7 ??!!
Parker et al. 2019, MNRAS accepted, arXiv:1811.10289v1 \\
Schlafly,  Meisner  \& Green, 2019, ApJS, 240, 30 \\
Simcoe et al., 2012, Nature, 492, 79 \\
Soares-Santos, et al. 2017, ApJ, 848, L16	\\
Stern et al. 2005, ApJ, 631, 163 \\
Stern et al. 2012,  ApJ, 753, 30 \\  %% MIR WISE Selection Paper I
Stern et al., 2018, ApJ, 864, 27 \\
Padovani et al. 2017, A\&ARv, 25, 2 \\
Rees, 1984, ARA\&A, 22, 471 \\
Rodr{\'{\i}}guez-Ardila \& Mazzalay, 2006, MNRAS, 367 L57 \\
Ross et al., 2018, MNRAS, 480, 4468 \\
Ross \& Cross, 2019, MNRAS, {\it in prep.} \href{https://github.com/d80b2t/VHzQ}{(draft available here)} \\
Valiante et al., 2018, MNRAS,  474, 3825 \\
van Velzen et al., 2016, ApJ, 829, 19 \\
Volonteri  et al., 2010, A\&ARv, 18, 279 \\
Wise et al., 2019, Nature,  arXiv:1901.07563v1 \\
Wright et al., 2010, AJ, 140, 1868 \\
Wyithe \& Loeb et al., 2003, ApJ, 586, 693 \\
Wang,~T., et al., 2018, MNRAS, 477, 2943 \\
Wang,~F., et al., 2018,  arXiv:1810.11926v1 \\
Yan et al., 2019, 1902.04163v1 \\
Yang et al,, 2018, arXiv:1811.11915v1 \\

%\end{multicols}

\end{document}